\documentclass[a4paper,prl,twocolumn,showpacs,superscriptaddress]{revtex4}%
\usepackage{amsfonts}
\usepackage{amsmath}
\usepackage{amssymb}
\usepackage{graphicx}%
\providecommand{\U}[1]{\protect\rule{.1in}{.1in}}

\begin{document}
\title{A general treatment of geometric phases and dynamical invariants}
\author{E. I. Duzzioni}
\email{duzzioni@ufabc.edu.br}
\affiliation{Centro de Ci\^{e}ncias Naturais e Humanas, Universidade Federal do ABC, Rua
Santa Ad\'{e}lia 166, Santo Andr\'{e}, S\~{a}o Paulo, 09210-170, Brazil}
\affiliation{Departamento de F\'{\i}sica, Universidade Federal de S\~{a}o Carlos,
13565-905, S\~{a}o Carlos, SP, Brazil}
\author{R. M. Serra}
\email{serra@ufabc.edu.br}
\affiliation{Centro de Ci\^{e}ncias Naturais e Humanas, Universidade Federal do ABC, Rua
Santa Ad\'{e}lia 166, Santo Andr\'{e}, S\~{a}o Paulo, 09210-170, Brazil}
\author{M. H. Y. Moussa}
\email{miled@ifsc.usp.br}
\affiliation{Instituto de F\'{\i}sica de S\~{a}o Carlos, Universidade de S\~{a}o Paulo,
Caixa Postal 369, S\~{a}o Carlos, S\~{a}o Paulo, 13560-970, Brazil}

\begin{abstract}
Based only on the parallel transport condition, we present a general method to
compute Abelian or non-Abelian geometric phases acquired by the basis states
of pure or mixed density operators, which also holds for nonadiabatic and
noncyclic evolution. Two interesting features of the non-Abelian geometric
phase obtained by our method stand out: i) it is a generalization of Wilczek
and Zee's non-Abelian holonomy, in that it describes nonadiabatic evolution
where the basis states are parallelly transported between distinct degenerate
subspaces, and ii) the non-Abelian character of our geometric phase relies on
the transitional evolution of the basis states, even in the nondegenerate
case. We apply our formalism to a two-level system evolving nonadiabatically
under spontaneous decay to emphasize the non-Abelian nature of the geometric
phase induced by the reservoir. We also show, through the generalized
invariant theory, that our general approach encompasses previous results in
the literature.

\end{abstract}

\pacs{03.65.Ca,03.65.Vf, 03.65.Yz}
\maketitle

\textit{Introduction}. The concept of geometric phase (GP) was transposed to
the domain of quantum systems undergoing cyclic adiabatic evolution by M.
Berry \cite{Berry 1984}, after having been introduced by Pancharatnam in
connection with interference of light waves with distinct polarizations
\cite{Pancharatnam 1956}. After Berry's discovery, Aharonov and Anandan
\cite{Aharonov 1987} removed the need for adiabatic evolution and Samuel and
Bhandari \cite{Samuel 1988} extended the concept of GP\textbf{ }to noncyclic
and nonunitary evolutions, introducing the notion of geodesic closure in the
projective Hilbert space. In recent years, the possibility of achieving
fault-tolerant quantum computation\ \cite{Zanardi 1999, Pachos 1999, Carollo
2005a} invoked the controversial subject of GPs for open quantum systems,
where the dynamic is generally nonunitary. In this case the GPs have been
defined by different methods: using phenomenological approaches \cite{Garrison
1988}, stochastic fields \cite{De Chiara 2003}, the usual master equation
\cite{Ellinas 1989}\ and quantum jumps \cite{Nazir 2002} techniques, apart
from state purification \cite{Tong 2004}, mean values of distributions
\cite{Marzlin 2004}, interferometric \cite{de Faria 2003}, and superoperator
methods \cite{Sarandy 2006, Goto 2007, Sarandy 2007}.

Parallel to these advances in the understanding of GPs, the dynamical
invariants (DIs) proposed by Lewis and Riesenfeld \cite{Lewis 1969} to handle
time-dependent Hamiltonians, have been applied to a number of problems,
including recent advances in cavity quantum electrodynamics \cite{Villas-Boas
2003} and Bose-Einstein condensates \cite{Duzzioni 2007}. Based on the DIs,
Morales \cite{Morales 1988} and Mizrahi \cite{Mizrahi 1989} introduced,
independently, a convenient way to compute the evolution of the GPs, which was
used to suggest an interferometric experiment to measure GPs induced by a
Stark shift in cavity quantum electrodynamics \cite{Duzzioni 2005}.

In this letter, we rely only on the parallel transport condition to obtain a
general formal expression to compute the GPs acquired by the basis states of a
density operator under unitary or nonunitary, adiabatic or nonadiabatic, and
cyclic or noncyclic evolutions. The DIs applied to the density matrix
\cite{Dodonov 1978} are required to account for the time-evolution of the GPs
in a general scenario of open quantum systems. Apart from showing that our
method reproduces previous results in literature, we use it to compute the GP
acquired by a two-level system under spontaneous decay. We show unambiguously
that this GP turns out to be non-Abelian even in nondegenerate system. Thus,
it is clear from this general approach that the non-Abelian character of the
GP is wholly associated with the transitional dynamics of the basis states.
Moreover, our treatment generalizes Wilczek and Zee's \cite{Wilczek 1984}
non-Abelian holonomy, in that it describes nonadiabatic evolutions in which
the basis states are parallelly transported between distinct degenerate subspaces.

\textit{Parallel transport condition}. To compute the GPs acquired by the
time-dependent states $\left\{  \left\vert \lambda,a;t\right\rangle \right\}
$ of an orthonormal basis, where $\lambda$ is $g$-fold degenerate
($a=1,2,...,g$), we first introduce the condition for parallel transport of a
state vector, namely that $\left.  _{\shortparallel}\left\langle
\lambda^{\prime},a^{\prime};t\right\vert d_{t}\left\vert \lambda
,a;t\right\rangle _{\shortparallel}\right.  =0$ \cite{Anandan 1988}. The
evolution operator responsible for the parallel transport during the time
interval $t$, $\left\vert \lambda,a;t\right\rangle _{\shortparallel
}=V_{\shortparallel}(t)\left\vert \lambda,a;0\right\rangle $, may be written
as $V_{\shortparallel}(t)=\sum_{\lambda,\lambda^{\prime}}\sum_{a,a^{\prime}%
}\left(  V_{\shortparallel}\right)  _{\lambda\lambda^{\prime}}^{aa^{\prime}%
}(t)\left\vert \lambda,a;t\right\rangle \left\langle \lambda^{\prime
},a^{\prime};0\right\vert $. Substituting $V_{\shortparallel}(t)$ in
the\ parallel transport condition, we find the equation for its coefficients
in the matrix form $\overset{\cdot}{V}_{\shortparallel}%
(t)=iA(t)V_{\shortparallel}(t)$, whose solution is given by%
\begin{equation}
V_{\shortparallel}(t)=T\exp\left(  i\int_{0}^{t}d\tau A(\tau)\right)  \text{,}
\label{Eq1}%
\end{equation}
where $V_{\shortparallel}(0)=\mathbf{1}$, the elements of the non-Abelian
connection $A(\tau)$ are $A_{\lambda,\lambda^{\prime}}^{a,a^{\prime}}%
(\tau)=i\left\langle \lambda,a;\tau\right\vert d_{\tau}\left\vert
\lambda^{\prime},a^{\prime};\tau\right\rangle $, and $T$ is the time-ordering
operator. The unitary matrix $V_{\shortparallel}(t)$ ($A(\tau)$ being
Hermitian) accounts for a cyclic and non-Abelian GP. As we are also concerned
with noncyclic evolutions, we must account for the relative phase acquired by
the parallelly transported state $\left\vert \lambda,a;t\right\rangle
_{\shortparallel}$ with respect to its starting point $\left\vert
\lambda,a;0\right\rangle $, given by $\arg$ $\left\{  \left\langle
\lambda,a;0\right\vert \left.  \lambda,a;t\right\rangle _{\shortparallel
}\right\}  $ $=\arg$ $\left\{  \left\langle \lambda,a;0\right\vert
W(t,0)V_{\shortparallel}(t)\left\vert \lambda,a;0\right\rangle \right\}  $,
with the elements of the \textit{overlap matrix} $W$ reading $W_{\lambda
,\lambda^{\prime}}^{a,a^{\prime}}(t,0)=\left\langle \lambda,a;0\right\vert
\left.  \lambda^{\prime},a^{\prime};t\right\rangle $. We note that, to
transport a subspace without rotating it locally, it must remain parallel to
itself during the time interval from $t$ to $t+\delta t$, i.e., $W_{\lambda
,\lambda^{\prime}}^{a,a^{\prime}}(t+\delta t,t)=\left\langle \lambda
,a;t\right\vert \left.  \lambda^{\prime},a^{\prime};t+\delta t\right\rangle
\simeq\delta_{\lambda,\lambda^{\prime}}\delta_{a,a^{\prime}}$. Consequently, a
necessary and sufficient condition stipulates that $A_{\lambda,\lambda
^{\prime}}^{a,a^{\prime}}(t)=0$, and hence $A(t)=0$.

As observed in Ref. \cite{Kult 2006}, in the general case, where the overlap
matrix $W(t,0)$ is restricted to an incomplete subspace $\left(
\sum\nolimits_{\lambda,a}\left\vert \lambda,a;t\right\rangle \left\langle
\lambda,a;t\right\vert \neq\mathbf{1}\right)  $, it must be decomposed in the
polar form $W=RU$, $R$ being a positive-definite (Hermitian) matrix ($\det
R>0$) and $U$ a unitary matrix ($\det U=\operatorname{e}^{i\varphi}$), such
that $\operatorname{e}^{i\varphi}=\det W/\det R$. In the case where $R$ is a
positive-semidefinite matrix ($\det R=0$), the Moore-Penrose pseudoinverse
must be used to evaluate the matrix $U$\textbf{. }Then, the non-Abelian,
nonadiabatic and noncyclic GP acquired by the basis states $\left\{
\left\vert \lambda,a;t\right\rangle \right\}  $ turns out to be
\begin{equation}
\mathcal{O}(t,0)=U(t,0)V_{\shortparallel}(t)\text{,} \label{Eq2}%
\end{equation}
where $U$ is the unitary part of $W$. Although Eq. (\ref{Eq2}) has already
been published by Mostafazadeh \cite{Mostafazadeh 1999} and Kult \textit{et
al}. \cite{Kult 2006}, in both cases the analyzed non-Abelian geometric phases
arise from the usual degeneracy, i.e., from transitions inside degenerate
subspaces. In a more general scenario, we have extended the analyses in Refs.
\cite{Mostafazadeh 1999,Kult 2006} to cases of transitional dynamics which
connect different degenerate and/or nondegenerate subspaces. As discussed
below, we found that transitional dynamics, even between nondegenerate states,
is the source of a non-Abelian geometric phase. Moreover, when considering the
case of transitional dynamics connecting different degenerate subspaces, our
analysis reveals a generalization of Wilczek and Zee's non-Abelian holonomy,
since it describes nonadiabatic evolutions where the basis states are
parallelly transported from one degenerate subspace to another. The definition
of parallel transport used here appear similar to that used in Ref.
\cite{Manini 2000}, however, our geometric quantity (\ref{Eq2}) is different
from that one defined in \cite{Manini 2000}. It is important to note that the
treatment presented here is general and includes several kinds of system dynamics.

The general expression for $\mathcal{O}(t,0)$ can be simplified in three
particular cases: transitional (t) dynamics in a degenerate (d) subspace
($\mathcal{O}^{\text{t,d}}(t,0)=\left[  U(t,0)V_{\shortparallel}(t)\right]
^{\text{t,d}}$) and transitional ($\mathcal{O}^{\text{t,nd}}(t,0)=\left[
U(t,0)V_{\shortparallel}(t)\right]  ^{\text{t,nd}}$) and non-transitional (nt)
($\mathcal{O}^{\text{nt,nd}}(t,0)=\left[  U(t,0)V_{\shortparallel}(t)\right]
^{\text{nt,nd}}$) dynamics in a nondegenerate (nd) subspace. The three
different cases mentioned above, covering all possible evolutions of the basis
states, will be explored below in the context of the DIs for a general
Lindblad evolution.

\textit{Gauge-invariance of the GP}. Under the gauge transformation
$\left\vert \lambda,a;t\right\rangle ^{\prime}=\sum\nolimits_{\nu,b}%
M_{\nu,\lambda}^{b,a}(t)\left\vert \nu,b;t\right\rangle $, where $M$ is a
unitary matrix, the operator $V_{\shortparallel}(t)$ becomes
$V_{\shortparallel}^{\prime}(t)=M^{\dagger}(t)V_{\shortparallel}(t)M(0)$, with
$A^{\prime}(t)=M^{\dagger}(t)A(t)M(t)+iM^{\dagger}(t)\overset{\cdot}{M}(t)$.
In its turn, $U(t,0)$ becomes $U^{\prime}(t,0)=M^{\dagger}(0)U(t,0)M(t)$. From
these transformed expressions, it follows that $\mathcal{O}^{\prime
}(t,0)=M^{\dagger}(0)\mathcal{O}(t,0)M(0)$, making the trace and the
eigenvalues of $\mathcal{O}^{\prime}(t,0)$ observable gauge invariants.

Summarizing our method, we first choose the time-dependent basis to expand the
state of the system $\rho(t)$. Next, we analyze the evolution of the basis
states to verify if there are transitions or not between them. With these
informations we build, through Eq. (2), the operators $\mathcal{O}%
^{\text{t,d}}(t,0)=\left[  U(t,0)V_{\shortparallel}(t)\right]  ^{\text{t,d}}$,
or $\mathcal{O}^{\text{t,nd}}(t,0)=\left[  U(t,0)V_{\shortparallel}(t)\right]
^{\text{t,nd}}$, or $\mathcal{O}^{\text{nt,nd}}(t,0)=\left[
U(t,0)V_{\shortparallel}(t)\right]  ^{\text{nt,nd}}$, given the geometric
phases acquired by the basis states under a general evolution.

\textit{GP and the DIs}. To illustrate the appearance of the GPs for the
different types of evolution discussed above, it is convenient to assume the
basis states $\left\{  \left\vert \lambda,a;t\right\rangle \right\}  $ to be
the eigenstates of an Hermitian DI, $I(t)\left\vert \lambda,a;t\right\rangle
=\lambda(t)\left\vert \lambda,a;t\right\rangle $, since the invariant method
is applicable for adiabatic and nonadiabatic processes under unitary and
nonunitary evolutions. Moreover, the information about open quantum dynamics
is naturally carried through the eigenstates of the invariant operators. In
fact, the DIs \cite{Lewis 1969} associate with any density operator $\rho(t)$
the time-conserved quantity $I(t)$, called invariant, which satisfies the
condition $\operatorname{d}\left\langle I(t)\right\rangle /\operatorname{d}%
t\equiv0$ \cite{Dodonov 1978}. From the Lindblad form for the density operator
of an open quantum system described by the Hamiltonian $H_{0}(t)$, given by
$\partial_{t}\rho(t)=-i\left[  H_{0}(t),\rho(t)\right]  +\sum\nolimits_{i,j}%
\gamma_{ij}(t)\left\{  \left[  \Gamma_{i},\rho(t)\Gamma_{j}^{\dag}\right]
+\left[  \Gamma_{i}\rho(t),\Gamma_{j}^{\dag}\right]  \right\}  $, the
evolution of the invariant operator following from $\left\langle
I(t)\right\rangle =\operatorname{Tr}\left[  I(t)\rho(t)\right]  $ becomes%
\begin{align}
\partial_{t}I(t)  &  =-i\left[  H_{0}(t),I(t)\right]  +\sum\nolimits_{i,j}%
\gamma_{ij}(t)\left\{  \Gamma_{j}^{\dag}\left[  \Gamma_{i},I(t)\right]
\right. \nonumber\\
&  \left.  +\left[  I(t),\Gamma_{j}^{\dag}\right]  \Gamma_{i}\right\}
\text{,} \label{Eq3}%
\end{align}
where $\Gamma_{i}$ are the Lindblad operators coming from the action of an
environment and $\gamma_{ij}(t)$ are the coupling strengths. We observe that
the Lindblad equation presented above applies for the case where the time
scale related to the rate of change of the time-dependent Hamiltonian
$H_{0}(t)$ is much slower than the time scales characterizing the reservoir
\cite{Davies 1978}.

\textit{Formal solution of the master equation and its connection with GPs.}
Expanding $\rho(t)$ in the eigenstate basis of the invariant,\ we obtain
through the master equation in the Lindblad form, the coupled differential
equations
\begin{gather}
\overset{\cdot}{c}_{\mu,\mu^{\prime}}^{b,b^{\prime}}=i%
{\displaystyle\sum\limits_{\lambda,a}}
\left(  H_{\mu,\lambda}^{b,a}+A_{\mu,\lambda}^{b,a}+iD_{\mu,\lambda}%
^{b,a}\right)  c_{\lambda,\mu^{\prime}}^{a,b^{\prime}}\nonumber\\
-i%
{\displaystyle\sum\limits_{\lambda,a}}
c_{\mu,\lambda}^{b,a}\left(  H_{\lambda,\mu^{\prime}}^{a,b^{\prime}%
}+A_{\lambda,\mu^{\prime}}^{a,b^{\prime}}-iD_{\lambda,\mu^{\prime}%
}^{a,b^{\prime}}\right) \nonumber\\
+2%
{\displaystyle\sum\limits_{i,j}}
\gamma_{ij}%
{\displaystyle\sum\limits_{\lambda,\lambda^{\prime},a,a^{\prime}}}
\Lambda_{\mu,\lambda;i}^{b,a}c_{\lambda,\lambda^{\prime}}^{a,a^{\prime}%
}\left(  \Lambda_{\mu^{\prime},\lambda^{\prime};j}^{b^{\prime},a^{\prime}%
}\right)  ^{\ast}\text{,} \label{Eq4}%
\end{gather}
where we have defined the time-dependent matrix elements $H_{\mu,\lambda
}^{b,a}=-\left\langle \mu,b;t\right\vert H_{0}\left\vert \lambda
,a;t\right\rangle $, $A_{\mu,\lambda}^{b,a}=\left\langle \mu,b;t\right\vert
id_{t}\left\vert \lambda,a;t\right\rangle $, $D_{\mu,\lambda}^{b,a}=$ $%
{\textstyle\sum\nolimits_{i,j}}
\gamma_{ij}\left\langle \mu,b;t\right\vert \Gamma_{j}^{\dag}\Gamma
_{i}\left\vert \lambda,a;t\right\rangle $, and $\Lambda_{\mu,\lambda;i}%
^{b,a}=\left\langle \mu,b;t\right\vert \Gamma_{i}\left\vert \lambda
,a;t\right\rangle $. It can be verified that the formal integration of Eq.
(\ref{Eq4}) leads to%
\begin{align}
c(t)  &  =T\exp\left\{  i\int\nolimits_{0}^{t}d\tau\left[  H+A+iD\right]
\left(  \cdot\right)  +2\sum\nolimits_{i,j}\int\nolimits_{0}^{t}d\tau\right.
\nonumber\\
&  \times\left.  \gamma_{ij}\Lambda_{i}\left(  \cdot\right)  \Lambda
_{j}^{\dagger}-i\left(  \cdot\right)  \int\nolimits_{0}^{t}d\tau\left[
H+A-iD\right]  \right\}  c(0)\text{,} \label{Eq5}%
\end{align}
with $\left(  \cdot\right)  $ indicating the side where the matrices $H+A\pm
iD$, $\Lambda_{i}$ and $\Lambda_{j}^{\dagger}$, are supposed to act\ on
$c(0)$, in the time-ordered expansion of the r.h.s. of (\ref{Eq5}).
Differently from the original procedure by Berry \cite{Berry 1984} (within the
context of\ an adiabatic and cyclic evolution of a pure state), it is no
longer clear, as first noted in Ref. \cite{Anandan 1988}, how to extract the
GP from the dynamics of the probability amplitudes (Eq. (\ref{Eq5})),
describing the nonadiabatic evolution of an open system, since in general $A$
does not commute with $H$, $D$, and $\Lambda_{i}$. Some exceptions correspond
to the specific cases of open quantum systems discussed in Refs. \cite{Sarandy
2006, Sarandy 2007} and filtering evolutions \cite{Kult 2007}. \textit{That is
why we have presented a formal definition of the GP in Eq. }(\ref{Eq2}%
)\textit{, instead of trying to obtain it, as usual }\cite{Berry 1984,Wilczek
1984,Mostafazadeh 1998,Mostafazadeh 1999}\textit{, from the dynamics of the
probability amplitudes}. For some particular situations of dissipative-free
dynamics, a connection can be established between the GPs emerging from our
definition and the dynamics of the probability amplitudes coefficients, as
shown below.

\textit{Dissipative-free dynamics (degenerate case)}. For dissipative-free
dynamics [$\gamma_{ij}(t)=0$ in Eq. (\ref{Eq3})] the solution (\ref{Eq5}),
inside the degenerate subspaces $\mu$ and $\mu^{\prime}$, reduces to%
\begin{align}
c_{\mu,\mu^{\prime}}(t)  &  =\left(  T\operatorname{e}^{i\int\nolimits_{0}%
^{t}d\tau\left(  H_{\mu,\mu}+A_{\mu,\mu}\right)  }\right)  c_{\mu,\mu^{\prime
}}(0)\nonumber\\
&  \times T\operatorname{e}^{-i\int\nolimits_{0}^{t}d\tau\left(
H_{\mu^{\prime},\mu^{\prime}}+A_{_{\mu^{\prime},\mu^{\prime}}}\right)
}\text{,} \label{Eq6}%
\end{align}
where $c_{\mu,\mu^{\prime}}$, $H_{\mu,\mu}$, and $A_{\mu,\mu}$ represent
matrices composed of the elements $c_{\mu,\mu^{\prime}}^{b,a}$, $H_{\mu,\mu
}^{b,a}$, and $A_{\mu,\mu}^{b,a}$. In Ref. \cite{Mostafazadeh 1998}, the
assumption that $\left[  H_{\mu,\mu}\text{, }A_{\mu,\mu}\right]  =0$ even for
a nonadiabatic evolution, resulted in the expressions $T\exp\left[
i\int\nolimits_{0}^{t}d\tau H_{\mu,\mu}(\tau)\right]  $ for dynamic and
$T\exp\left[  i\int\nolimits_{0}^{t}d\tau A_{\mu,\mu}(\tau)\right]  $ for
cyclic geometric phases. However, as in general $\left[  H_{\mu,\mu}\text{,
}A_{\mu,\mu}\right]  \neq0$ for transitional and nonadiabatic dynamics in a
degenerate subspace $\mu$, we must return to the formal expression (\ref{Eq2})
with $V_{\shortparallel}^{\text{t,d}}(t)=\sum_{\mu}\sum_{a,a^{\prime}}\left(
V_{\shortparallel}\right)  _{\mu\mu}^{aa^{\prime}}(t)\left\vert \mu
,a;t\right\rangle \left\langle \mu,a^{\prime};0\right\vert $ to obtain the GPs
for cyclic ($U=\mathbf{1}$) and noncyclic ($U\neq\mathbf{1}$) evolutions. For
the cyclic case, Eq. (\ref{Eq2}) with $V_{\shortparallel}^{\text{t,d}}(t)$
leads exactly to the expression $T\exp\left[  i\int\nolimits_{0}^{t}d\tau
A_{\mu,\mu}(\tau)\right]  $ \cite{Mostafazadeh 1998} for the GP, whereas for
the noncyclic case the matrix $U$ must be taken into account (instead of
matrix $W$ as in Ref. \cite{Mostafazadeh 1999}), with the elements of the
non-Abelian connection being given by $A_{\mu,\mu}^{a,a^{\prime}}%
(\tau)=i\left\langle \mu,a;\tau\right\vert d_{\tau}\left\vert \mu,a^{\prime
};\tau\right\rangle $.

For transitional and adiabatic dynamics in a degenerate subspace, the
adiabatic evolution $V_{\shortparallel}^{\text{t,d}}(t)$ is built up from the
condition $\partial_{t}I\simeq0$ \cite{Mizrahi 1989}, which implies that
$H_{0}$ and $I$ assume the same basis $\left\{  \left\vert \mu
,a;t\right\rangle \right\}  $. In this case, the non-Abelian dynamic phase
reduces to the Abelian one, since $H_{\mu,\mu}^{b,a}=E_{\mu}\delta_{ab}$,
where $E_{\mu}$ is the eigenenergy associated with the state $\left\vert
\mu,a;t\right\rangle $ and, consequently, $\left[  H_{\mu,\mu}\text{, }%
A_{\mu,\mu}\right]  =0$. The cyclic GP thus remains the non-Abelian holonomy
$T\exp\left[  i\int\nolimits_{0}^{t}d\tau A(\tau)\right]  $ computed in the
Hamiltonian eigenstates, obtained by Wilczek and Zee \cite{Wilczek 1984}. For
a noncyclic evolution, the nonidentity matrix $U$ is responsible for yielding
the result in Ref. \cite{Kult 2006}.

\textit{Dissipative-free dynamics (nondegenerate case). }For the nondegenerate
case, we easily verify that the solution of (\ref{Eq4}), corresponding to a
particular case of Eq. (\ref{Eq6}), is given by $c_{\mu,\mu^{\prime}%
}(t)=c_{\mu,\mu^{\prime}}(0)\exp\left\{  i\int\nolimits_{0}^{t}d\tau\left[
\Delta H_{\mu,\mu^{\prime}}(\tau)+\Delta A_{\mu,\mu^{\prime}}(\tau)\right]
\right\}  $, with $\Delta H_{\mu,\mu^{\prime}}=H_{\mu,\mu}-H_{\mu^{\prime}%
,\mu^{\prime}}$, $H_{\mu,\mu}=-\left\langle \mu;t\right\vert H_{0}\left\vert
\mu;t\right\rangle $, $\Delta A_{\mu,\mu^{\prime}}=A_{\mu,\mu}-A_{\mu^{\prime
},\mu^{\prime}}$, and $A_{\mu,\mu}=\left\langle \mu;t\right\vert
id_{t}\left\vert \mu;t\right\rangle $, so that the dynamic and cyclic GPs
reduce to the Abelian expressions $\exp\left[  i\int\nolimits_{0}^{t}d\tau
H_{\mu,\mu}(\tau)\right]  $ and $\exp\left[  i\int\nolimits_{0}^{t}d\tau
A_{\mu,\mu}(\tau)\right]  $. Note that the dynamic and geometric phases
associated with the element $\rho_{\mu,\mu^{\prime}}$ is simply the difference
between the phases acquired by the states $\left\vert \mu;t\right\rangle $ and
$\left\vert \mu^{\prime};t\right\rangle $, even for the nonadiabatic case, as
can be verified from the DIs \cite{Lewis 1969}. This striking feature of the
DIs enable nonadiabatic evolutions with non-transitional eigenstates in the
nondegenerate case. In this connection, the operators $U^{\text{nt,nd}}%
(t,0)$\textbf{ }and\textbf{ }$V_{\shortparallel}^{\text{nt,nd}}(t)=\sum
_{\lambda}\left(  V_{\shortparallel}\right)  _{\lambda\lambda}(t)\left\vert
\lambda;t\right\rangle \left\langle \lambda;0\right\vert $\textbf{ }in Eq.
(\ref{Eq2})\textbf{ }describe the GP for noncyclic\textbf{ }evolutions whose
expression is $\left\langle \lambda;0\right.  \left\vert \lambda
;t\right\rangle \exp\left(  -\int\nolimits_{0}^{t}d\tau\left\langle
\lambda;\tau\right\vert d_{\tau}\left\vert \lambda;\tau\right\rangle \right)
$ \cite{Mostafazadeh 1999}. For cyclic evolutions this GP reduces to that
obtained in Refs. \cite{Morales 1988, Mizrahi 1989}. Evidently, for the
adiabatic case ($\partial_{t}I\simeq0$) the above expression for the noncyclic
GP gives the result in Ref. \cite{Polavieja 1998}, while in the cyclic case it
reproduces the original Berry phase \cite{Berry 1984}.

\textit{Application - An unstable two-level system}. A particular form of the
operator $V_{\shortparallel}(t)$, given by $\left[  V_{\shortparallel
}(t)\right]  ^{\text{t,nd}}=\sum_{\lambda,\lambda^{\prime}}\left(
V_{\shortparallel}\right)  _{\lambda\lambda^{\prime}}(t)\left\vert
\lambda;t\right\rangle \left\langle \lambda^{\prime};0\right\vert $, occurs
when we have a nonadiabatic evolution with a transition between non-degenerate
states of the basis $\left\{  \left\vert \lambda;t\right\rangle \right\}  $.
This is the case in a quantum system undergoing a unitary evolution, as in
Ref. \cite{Anandan 1988}, or a nonunitary evolution where, as shown below, the
noise injection gives rise to a non-Abelian GP.

Employing the above method to analyze the role of dissipation in the evolution
of the GP, we consider a non-degenerate two-level system, with transition
frequency $\omega_{0}$ between the ground ($g$) and excited ($e$) states,
under spontaneous decay at 0K. The dynamics of this system is described by the
solution to the master equation for $\rho(t)$ with $H_{0}=\omega_{0}\sigma
_{z}/2$, $\gamma_{11}=\gamma/2$, and $\Gamma_{1}=\sigma_{-}$. For the
invariant operator, we assume
\begin{equation}
I(t)=\sum\nolimits_{\alpha,\beta=g,e}\chi_{\alpha\beta}(t)\sigma_{\alpha\beta
}\text{,} \label{7}%
\end{equation}
where $\sigma_{\alpha\beta}=\left\vert \alpha\right\rangle \left\langle
\beta\right\vert $ and the coefficients $\chi_{\alpha\beta}(t)$, which are
solutions of Eq. (\ref{Eq3}), satisfy the relations
\begin{subequations}
\label{8}%
\begin{align}
\chi_{gg}(t)  &  =-r_{0}\cos\theta_{0}\text{,}\label{8a}\\
\chi_{ee}(t)  &  =\left(  2\operatorname{e}^{\gamma t}-1\right)  r_{0}%
\cos\theta_{0}\text{,}\label{8b}\\
\chi_{eg}(t)  &  =r_{0}\sin\theta_{0}\operatorname{e}^{\gamma t/2-i\left(
\omega_{0}t+\phi_{0}\right)  }\text{,}\label{8c}\\
\chi_{ge}(t)  &  =r_{0}\sin\theta_{0}\operatorname{e}^{\gamma t/2+i\left(
\omega_{0}t+\phi_{0}\right)  }\text{,} \label{8d}%
\end{align}
with $\theta_{0}$ ($\phi_{0}$) being the polar (azimuthal) angle of the
initial pure ($r_{0}=1$) or mixed ($r_{0}<1$) state $\rho(0)$ in the Bloch
sphere. The eigenvalues and eigenvectors of the invariant (\ref{7}) are given
by $\lambda_{\pm}(t)=-r_{0}\left[  \left(  1-\operatorname{e}^{\gamma
t}\right)  \cos\theta_{0}\mp\operatorname{e}^{\gamma t/2}\sqrt{1-\left(
1-\operatorname{e}^{\gamma t}\right)  \cos^{2}\theta_{0}}\right]  $,
$\left\vert \pm;t\right\rangle $ $=\pm N(t)\left\{  f(t)\left\vert
\genfrac{}{}{0pt}{}{g}{e}%
\right\rangle \pm r_{0}\sin\theta_{0}\operatorname{e}^{\mp i\left(  \omega
_{0}t+\phi_{0}\right)  }\left\vert
\genfrac{}{}{0pt}{}{e}{g}%
\right\rangle \right\}  $, with $f(t)=-r_{0}\left[  \operatorname{e}^{\gamma
t/2}\cos\theta_{0}-\sqrt{1-\left(  1-\operatorname{e}^{\gamma t}\right)
\cos^{2}\theta_{0}}\right]  $ and $N^{2}(t)=$ $\left(  f^{2}(t)+r_{0}^{2}%
\sin^{2}\theta_{0}\right)  ^{-1}$.

In order to find the coefficients $\left(  V_{\shortparallel}\right)
_{\lambda\lambda^{\prime}}(t)$ of $\left[  V_{\shortparallel}(t)\right]
^{\text{t,nd}}$ we first solve the equation $\left[  \overset{\cdot}%
{V}_{\shortparallel}(t)\right]  ^{\text{t,nd}}=iA(t)\left[  V_{\shortparallel
}(t)\right]  ^{\text{t,nd}}$, where the elements of the non-Abelian connection
are given by $A_{kl}=i\left\langle k;t\right\vert d_{t}\left\vert
l;t\right\rangle $, with $k,l=+,-$. To this end, we move to the rotating frame
$R=\exp\left[  \eta\left(  \operatorname{e}^{-i\zeta}\sigma_{-}%
-\operatorname{e}^{i\zeta}\sigma_{+}\right)  /2\right]  $, as in Ref.
\cite{Duzzioni 2007}, obtaining the coupled differential equations
\end{subequations}
\begin{subequations}
\label{10}%
\begin{align}
\dot{\eta}  &  =2N^{2}r_{0}\sin\theta_{0}\left[  \omega_{0}f\sin\Theta+\dot
{f}\cos\Theta\right]  \text{,}\label{10a}\\
\dot{\zeta}  &  =2N^{2}r_{0}\sin\theta_{0}\left\{  \omega_{0}r_{0}\sin
\theta_{0}\right. \nonumber\\
&  \left.  +\cot\eta\left[  -\omega_{0}f\cos\Theta+\dot{f}\sin\Theta\right]
\right\}  \text{,} \label{10b}%
\end{align}
where $\Theta=\omega_{0}t+\phi_{0}-\zeta$. Under the initial condition
$\pi/7\lesssim\theta_{0}\lesssim\pi$ \textbf{(}which is necessary to obtain an
approximated analytical solution\textbf{)}, the assumption of a typical weak
system-reservoir coupling, $\gamma/\omega_{0}\ll1$, and assuming a time
evolution around $t\simeq2\pi/\omega_{0}$,\ we obtain $\dot{\eta}\simeq0$,
such that $\eta(t)\simeq\operatorname{arccot}\left\{  \cot\theta_{0}\left[
1+\frac{\left(  1-\cos\theta_{0}/2\right)  }{1-\cos\theta_{0}}\gamma t\right]
\right\}  $ and $\zeta(t)=\omega_{0}t+\phi_{0}-\gamma\cos\theta_{0}%
/2\omega_{0}$.

>From the above result, we obtain the noncyclic, nonadiabatic, and non-Abelian
GP associated with the decaying two-level system $\mathcal{O}^{\text{t,nd}%
}(t,0)$ $=U(t,0)R^{\dagger}(t)\operatorname{e}^{i\Omega(t)\sigma_{z}}R(0)$,
where $\Omega(t)\simeq\omega_{0}t\left[  1+\gamma S(\theta_{0})t/2\right]  $,
$S(\theta_{0})=-\cos\theta_{0}\left(  1/2-\cos\theta_{0}+3\cos^{2}\theta
_{0}/8\right)  /(1-\cos\theta_{0})$, and
\end{subequations}
\begin{equation}
U(t,0)=\left(
\begin{array}
[c]{cc}%
U_{D} & -U_{OD}^{\ast}\\
U_{OD} & U_{D}^{\ast}%
\end{array}
\right)  \text{,} \label{13}%
\end{equation}
with the on- and off-diagonal elements of the overlap matrix given,
respectively, by $U_{D}=N\sin\left(  \theta_{0}/2\right)  \left[
f+2r_{0}\operatorname{e}^{-i\omega_{0}t}\cos^{2}\left(  \theta_{0}/2\right)
\right]  $ and $U_{OD}=N\operatorname{e}^{-i\phi_{0}}\cos\left(  \theta
_{0}/2\right)  \left[  f-2r_{0}\operatorname{e}^{-i\omega_{0}t}\sin^{2}\left(
\theta_{0}/2\right)  \right]  $. At this point, we stress that the decay
process introduced by the reservoir leads to a transitional dynamics of the DI
basis states, bringing about a not fault-tolerant non-Abelian GP. In the
particular case $\gamma=0$ (nondegenerate dissipative-free dynamics), the
\textit{non-transitional evolution} during the time interval $2\pi/\omega_{0}$
leads to the cyclic GPs $\mathcal{O}_{\pm}^{\text{nt,nd}}(2\pi/\omega
_{0},0)=\pm\pi(1-\cos\theta_{0})$ associated with the eigenstates $\left\vert
\pm;t\right\rangle $, as obtained in Refs. (\cite{Berry 1984,Mizrahi 1989}).

Whereas some works are concerned with a formal definition of the GPs for open
quantum systems \cite{Tong 2004, Marzlin 2004, de Faria 2003, Sarandy 2006,
Goto 2007, Sarandy 2007}, most of them restrict themselves to compute
corrections to this phase coming from the reservoir \cite{Garrison 1988, De
Chiara 2003, Ellinas 1989, Nazir 2002}. From the formal approach presented in
this work we additionally verify that the reservoir may even change the nature
of the holonomy, from an Abelian $\left(  \gamma=0\right)  $ to a non-Abelian
$\left(  \gamma\neq0\right)  $ one, as it emerges from the above application.

Summarizing, we have presented a general formalism to compute GPs, starting
only from the parallel transport condition. These GPs transform covariantly
and the approach is applicable to a general scenario, including adiabatic or
nonadiabatic, cyclic or noncyclic, and transitional or non-transitional
evolutions of pure or mixed states. Although we have used the DIs to compute
the GPs acquired by the basis states of the invariant, the formalism is
applicable to any time-dependent basis states. Besides reproducing well-known
results established in the literature, our formal definition reveals two
striking features of the GP: it generalizes Wilczek and Zee's non-Abelian
holonomy \cite{Wilczek 1984}, in describing nonadiabatic evolutions where the
basis states are parallelly transported between distinct degenerate subspaces;
secondly, our method demonstrates clearly that the non-Abelian character of
the GP arises from transitional dynamics, even in nondegenerate case. We have
shown that the nonadiabatic evolution of an open two-level quantum system
introduces a non-Abelian holonomy. Both of these features not only deepen our
understanding of GPs, but also offer the possibility of investigating how to
use the non-Abelian holonomy acquired by transitional dynamics between
nondegenerate states to perform geometric quantum computation.

\begin{acknowledgments}
We gratefully acknowledge financial support from the Brazilian agencies CNPq
and UFABC (to E.I.D.), and CNPq and FAPESP (to R.M.S. and M.H.Y.M.). We are
also grateful to an anonymous referee for valuable comments.
\end{acknowledgments}


\begin{thebibliography}{99}                                                                                               %


\bibitem {Berry 1984}M. V. Berry, Proc. Roy. Soc. London A \textbf{392}, 45 (1984).

\bibitem {Pancharatnam 1956}S. Pancharatnam, Proc. Ind. Acad. Sci.
\textbf{A44}, 247 (1956).

\bibitem {Aharonov 1987}Y. Aharonov and J. Anandan, Phys. Rev. Lett.
\textbf{58}, 1593 (1987).

\bibitem {Samuel 1988}J. Samuel and R. Bhandari, Phys. Rev. Lett. \textbf{60},
2339 (1988).

\bibitem {Zanardi 1999}P. Zanardi and M. Rasetti, Phys. Lett. A \textbf{264}%
,\textbf{\ }94 (1999).

\bibitem {Pachos 1999}J. Pachos, \textit{et al}., Phys. Rev. A \textbf{61}%
,\textbf{\ }010305(R) (1999); A. Recati, \textit{et al}., \textit{ibid}.
\textbf{66},\textbf{\ }032309 (2002); P. Solinas, \textit{et al}.,
\textit{ibid}. \textbf{67}, 052309 (2003); Li-X. Cen and P. Zanardi,
\textit{ibid}. \textbf{70}, 052323 (2004); A. Ekert, \textit{et al}., J. Mod.
Opt. \textbf{47}, 2501 (2000);\ L. M. Duan, \textit{et al}., Science
\textbf{292},\textbf{\ }1695 (2001); L. Faoro, J. Siewert, and R. Fazio, Phys.
Rev. Lett. \textbf{90}, 028301 (2003); I. Fuentes-Guridi, \textit{et al}.,
\textit{ibid}. \textbf{94}, 020503 (2005).

\bibitem {Carollo 2005a}A. Carollo and V. Vedral, quant-ph/0504205 (2005).

\bibitem {Garrison 1988}J. C. Garrison and E. M. Wright, Phys. Lett. A
\textbf{128}, 177 (1988); G. Dattoli, R. Mignani, and A. Torre, J. Phys. A
\textbf{2}3, 5795 (1990).

\bibitem {De Chiara 2003}G. De Chiara and G. M. Palma, Phys. Rev. Lett.
\textbf{91}, 090404 (2003); A. Blais and A. --M. S. Tremblay, Phys. Rev. A
\textbf{67}, 012308 (2003).

\bibitem {Ellinas 1989}D. Ellinas, S. M. Barnett, and M. A. Dupertuis, Phys.
Rev. A \textbf{39}, 3228 (1989); D. Gamliel and J. H. Freed, Phys. Rev. A
\textbf{39}, 3238 (1989); K. M. F. Romero, A. C. A. Pinto, and M. T. Thomaz,
Physica A \textbf{307}, 142 (2002); A. C. A. Pinto and M. T. Thomaz, J. Phys.
A \textbf{36}, 7461 (2003); R. S. Whitney and Y. Gefen, Phys. Rev. Lett.
\textbf{90}, 190402 (2003); I. Kamleitner, J. D. Cresser, and B. C. Sanders,
Phys. Rev. A \textbf{70}, 044103 (2004); R. S. Whitney, Y. Makhlin, A.
Shnirman, and Y. Gefen, Phys. Rev. Lett. \textbf{94}, 070407 (2005); F. C.
Lombardo and P. I. Villar, Phys. Rev. A \textbf{74}, 042311 (2006).

\bibitem {Nazir 2002}A. Nazir, T. P. Spiller, and W. J. Munro, Phys. Rev. A
\textbf{65}, 042303 (2002); A. Carollo, \textit{et al.}, Phys. Rev. Lett.
\textbf{90}, 160402 (2003); A. Bassi and E. Ippoliti, Phys. Rev. A
\textbf{73}, 062104 (2006).

\bibitem {Tong 2004}D. M. Tong, \textit{et al.}, Phys. Rev. Lett. \textbf{93},
080405 (2004); D. M. Tong, \textit{et al}., \textit{ibid}. \textbf{95},
249902(E) (2005); A. T. Rezakhani and P. Zanardi, Phys. Rev. A \textbf{73},
012107 (2006).

\bibitem {Marzlin 2004}K. -P. Marzlin, \textit{et al}., Phys. Rev. Lett.
\textbf{93}, 260402 (2004).

\bibitem {de Faria 2003}J. G. P. de Faria, A. F. R. de Toledo Piza, and M. C.
Nemes, Europhys. Lett. \textbf{62}, 782 (2003); M. Ericsson, \textit{et al.},
Phys. Rev. A \textbf{67}, 020101(R) (2003).

\bibitem {Sarandy 2006}M. S. Sarandy and D. A. Lidar, Phys. Rev. A
\textbf{73}, 062101 (2006).

\bibitem {Goto 2007}H. Goto and K. Ichimura, Phys. Rev. A \textbf{76}, 012120 (2007).

\bibitem {Sarandy 2007}M. S. Sarandy, E. I. Duzzioni, and M. H. Y. Moussa,
Phys. Rev. A \textbf{76}, 052112 (2007).

\bibitem {Lewis 1969}H. R. Lewis and W. B. Riesenfeld, J. Math. Phys.
\textbf{10}, 1458 (1969).

\bibitem {Villas-Boas 2003}C. J. Villas-Boas, \textit{et al}., Phys. Rev.
\textbf{68}, 053808 (2003).

\bibitem {Duzzioni 2007}E. I. Duzzioni, \textit{et al.}, Phys. Rev. A
\textbf{75}, 032113 (2007).

\bibitem {Morales 1988}D. A. Morales, J. Phys. A: Math. Gen. \textbf{21}, L889 (1988).

\bibitem {Mizrahi 1989}S. S. Mizrahi, Phys. Lett. A \textbf{138}, 465 (1989).

\bibitem {Duzzioni 2005}E. I. Duzzioni, \textit{et al.}, Europhysics Letters
\textbf{72}, 21 (2005).

\bibitem {Dodonov 1978}V. V. Dodonov and V. I. Man'ko, Physica A \textbf{94,
}403 (1978).

\bibitem {Wilczek 1984}F. Wilczek and A. Zee, Phys. Rev. Lett. \textbf{52},
2111 (1984).

\bibitem {Anandan 1988}J. Anandan, Phys. Lett. A \textbf{133}, 171 (1988).

\bibitem {Kult 2006}D. Kult, \textit{et al}., Phys. Rev. \textbf{74}, 022106 (2006).

\bibitem {Mostafazadeh 1999}A. Mostafazadeh, J. Phys. A: Math. Gen.
\textbf{32}, 8157 (1999).

\bibitem {Manini 2000}N. Manini and F. Pistolesi, Phys. Rev. Lett.
\textbf{85}, 3067 (2000).

\bibitem {Davies 1978}E. B. Davies and H. Spohn, J. Stat. Phys. \textbf{19},
511 (1978); G. Florio, \textit{et al.}, Phys. Rev. A \textbf{73}, 022327 (2006).

\bibitem {Kult 2007}D. Kult, J. Aberg, and E. Sj\"{o}qvist, Europhys. Lett.
\textbf{78}, 60004 (2007).

\bibitem {Mostafazadeh 1998}A. Mostafazadeh, J. Phys. A: Math. Gen.
\textbf{31}, 9975 (1998).

\bibitem {Polavieja 1998}G. G. de Polavieja and E. Sj\"{o}qvist, Am. J. Phys.
\textbf{66}, 431 (1998).
\end{thebibliography}
\end{document}